\documentclass[twoside,twocolumn,headings=normal,titlepage=false,]{scrartcl}

\usepackage{./style_CEAA}

\title{Compositional Equivalence with actor attributes: Positional analysis of the the Florentine Families network} 
\author{J. Antonio Rivero Ostoic\textsuperscript{a}\thanks{\textit{Email address:} \texttt{rivero.antonio@gmail.com} \protect\\ \!\!\textit{URL address:} \texttt{https://github.com/mplex}} 
        \protect\\[5pt] \small  \textsuperscript{a}\textit{CESU, University of San Simón, Cochabamba, Bolivia}}
\date{}
\publishers{%
\normalsize
\vspace{0.5cm} 
\rule{\textwidth}{1pt}\bigbreak
\parbox{0.9\linewidth}{\sl
This paper extends Compositional Equivalence ---which is a structural correspondence type aimed for multiplex networks--- by incorporating actor attributes in the modelling of the network relational structure as diagonal matrices. As an illustration, we construct the positional system of the Florentine families' network in the 15th century with Business and Marriage ties together with relevant characteristics acquired from the actors such as families' financial Wealth and their number of Priorates. Different representations of the cumulated person hierarchies reveal that adding Wealth in the modelling provides a more accurate picture of what the substantial narrative says about this network.
}
\bigbreak\rule{\textwidth}{1pt}
}

\setkomafont{disposition}{\bfseries}
\usepackage[manualmark,headsepline]{scrlayer-scrpage} 
\begin{document}

\maketitle

\newpage

\singlespacing

\section{Introduction}
\noindent
Relationships among actors in a defined collective scheme are the primary source of information for social network analysis. Ties not only make the network structure, but they also provide the basis for the characterisation of underlying processes occurring in the social system. Although social networks are typically characterised by a single type of relationship, social life is more complex and people are embedded with ``different'' kinds of ties that are interlocked within the network relational structure. 

These sorts of arrangements are known as \emph{multiplex networks}, and the associated relational structure of such social systems is typically reduced onto positional systems to facilitate a useful substantial interpretation. A key aspect in the reduction process is to preserve the multiplicity of the ties since the way different ties are intertwined provides important information about the network structure. 

In this spirit, \citet{BrePat86} proposed a type of equivalence among the network members that is built on local role algebras for the creation of the positional system. Our goal with this paper is to extend this type of correspondence with an effective way to incorporate the attributes of the actors and their relationships into a single relational system representing the multiple network structure. One important reason for such integration is that social conduct in networks does not always institute a link between individual subjects, and attribute-based information the actors is often not ascribed to them, but depends on the individual's own choices or circumstances. 

Examples of actor attributes are the acquisition of a certain characteristic from the social environment such as innovation adoption, the taking of a certain attitude, the non-compulsory affiliation to a group, the people's personal wealth and political power, etc. Such attributes can play a significant role in the network relational structure and they should be incorporated in the modelling process.

\section{Algebra for Multiple Networks}
\noindent
Representing social relations and actor attributes in an integrated system requires a formal definition of the social network concept. A \emph{social network} $X$ comprises a set $N$ of $n$ social actors, $N = \{ i \mid i \text{ is an actor} \}$ measured under a collection of social relations $R = \{ ( i, j ) \mid i \;\text{ `has a tie to'} j \}$ being $(i,j)$ an ordered pair. A binary relation $X_R(i,j)=1$ represents a tie $R$ between actors $i$ and $j$ in $X$, whereas $X_R(i,j)=0$ denotes the lack of a tie. The pairs on $X_R$ are stored in an adjacency matrix $A$ with size $n \times n$. 

A multiple network $\Xx$ is a collection $\Rr$ of $r$ different kinds of relations, $\Rr=\{R_1, R_2, \dotss, R_r\}$ measured over $N$. In this case each relational type is stored in separate adjacency matrices $A_1, A_2, \dotss, A_r$, which are stacked together into a single array $\Aa$ with size $n \times n \times r$. Moreover, the actors and their ties are also represented by nodes and differentiated edges, respectively, in a graphical device called a \emph{multigraph}, in which the relational levels are depicted in parallel rather than being collapsed into bold edges representing multiplex ties.

Each element in $\Rr$ constitutes a generator tie that produces \emph{compound} relationships among the network members through relational composition, and compound ties can be concatenated as well. For instance, the ``the friend of a colleague'' comes from generators ``friend of'' and ``colleague of'', etc. Both generators and compounds are referred as \emph{strings} in relational structures.

\subsection{Representation of Attributes}
\noindent
One of the theses of this paper is that non-ascribed attributes from the actors in the network can be an integrated part of the relational structure, which is typically represented by a semigroup of relations \citep{BoorWhite76, Pattison1993}. In this sense, the incorporation of the changing attributes of the actors implies that subjects sharing a characteristic constitute a subset of self-reflexive ties associated to the social system represented with a matrix format to be combined with the other elements in the relational structure.

In formal terms, actor attributes are to be represented by the elements of an \emph{diagonal matrix} $A^{\alpha}$ where each value is defined as:
$$
a_{ij}^{\alpha} \;=\; c_i \delta_{ij}.
$$
\nov
Accordingly, for a given attribute defined in $\alpha$, and for $i = x_1, x_2, \dotss, x_n$, the possible values of the first variable in the right hand expression are:
$$
c_i =
\begin{cases}
	1 \quad \text{if the attribute is tied to actor } i  \\  
	0 \quad \text{otherwise.}  
\end{cases}
$$
\nov
On the other hand, $\delta_{ij}$ is defined for nodes $i, j = x_1, x_2, \dotss, x_n$ in $\Xx$ by the delta function or Kronecker delta as:
$$
\delta_{ij} =
\begin{cases}
	1 \quad \text{for }\; i = j  \\
	0 \quad \text{for }\; i \ne j.
\end{cases}
$$
\nov
As a result, the general representation of $A^\alpha$ constitutes a diagonal matrix with the form:
$$
 \begin{pmatrix}
  c_1   &    0   & \dots  &   0    \\
   0    &    c_2 & \dots  &   0    \\
 \vdots & \vdots & \ddots & \vdots \\
   0    &    0   & \dots  &  c_n   \\
 \end{pmatrix}
$$
\nov
that records as self-relationships the attributes of the total number of actors in the system. In other words, the `possession' of the attribute produces a reflexive closure in the respective element of the system.

The establishment of the indexed diagonal matrix implies that each type of attribute considered for the actors in the network is represented by its own array, and it constitutes an additional generator in the relational structure. In case all network members share a given attribute, the result will be an identity matrix without any structural effect, whereas in case none of the actors possesses the characteristic, the representation will be a null matrix with an annihilating effect where no composition is possible. 

Clearly, we are mainly interested in the differentiation of the actors who share an attribute as opposed to those who do not share the trait because the resulting matrix that is neither a neutral nor an absorbing element in the algebraic structure has structuring consequences in the network relational system.

\section{Equivalence in Multiple Networks}
\noindent
Although the concatenation of social ties used in the construction of the partial order structure is well assumed \citep{BoorWhite76, Pattison1993}, there are caveats in producing algebraic systems with the diagonal matrices representing actor attributes. For instance, since social interactions are typically measured without loops and are represented by adjacency matrices with empty diagonals, these cannot be contained in an attribute relation with this form of representation. However, by grouping actors who are structurally equivalent, it is possible to obtain collective self-relations.

Take relations $\cC$ and $\aA$ (and $\fF$) below for example. Relation $\cC$ has three maximally connected actors that make a clique configuration, but just a couple of them share attribute $\aA$. Intrinsically, this means that the network of $\cC$ relations includes system $\aA$ and certainly not the other way around, but in the given example there is not such containment and it does not reflect the reality of this network. 

\begin{center}
\renewcommand{\arraystretch}{.8}
\begin{tabular}{ccc}
\begin{tabular}{>{$}c<{$} >{$}c<{$} >{$}c<{$} }
	 0 & 1 & 1 \\
	 1 & 0 & 1 \\
	 1 & 1 & 0 \\
\end{tabular}
&
\begin{tabular}{>{$}c<{$} >{$}c<{$} >{$}c<{$} }
	 0 & 0 & 0 \\
	 0 & 0 & 0 \\
	 1 & 1 & 0 \\
\end{tabular}
&
\begin{tabular}{>{$}c<{$} >{$}c<{$} >{$}c<{$} }
	 1 & 0 & 0 \\
	 0 & 1 & 0 \\
	 0 & 0 & 0 \\
\end{tabular} \\[15pt]
$\cC$ & $\fF$ & $\aA$ \\
\end{tabular}
\end{center}

\nov
A solution to this issue is to group the structurally equivalent actors in the network, which is to the same as categorising actors with similar patterns of relationships. For instance, it is obvious that actors $1$ and $2$ are structural equivalent according to the definition made by \citet{LorrWht71} because these actors are identically related with both social relations ($\cC$ and $\fF$), and they even share the same attribute. As a result, the first two actors make up a single class and the associations in the system now echo the inclusion $\aA \leq \cC$, thanks to the reflexive character of the first class of actors.

\begin{center}
\renewcommand{\arraystretch}{.8}
\begin{tabular}{ccc}
\begin{tabular}{>{$}c<{$} >{$}c<{$} }
	 1 & 1  \\
	 1 & 0  \\
\end{tabular}
&
\begin{tabular}{>{$}c<{$} >{$}c<{$} }
	 0 & 0  \\
	 1 & 0  \\
\end{tabular}
&
\begin{tabular}{>{$}c<{$} >{$}c<{$} }
	 1 & 0 \\
	 0 & 0 \\
\end{tabular} \\[10pt]
$\cC$ & $\fF$ & $\aA$ \\
\end{tabular}
\end{center}

\nov
Structural Equivalence is the most stringent type of correspondence and since its formal definition significant relaxations have been proposed for social networks; notably Automorphic Equivalence \citep{WinMan83, Everett85}, Regular Equivalence \citep{WhteRtz83, Sailer78}, and Generalized Equivalence \citep{DorBatFer94, DorBatFer2004}. A common characteristic among these correspondence types is that they have a \emph{global} perspective because the standpoints of the entire set of actors within the social system are taken into account simultaneously in the modelling process. Another distinctive feature of these equivalences is that they were originally designed for simple networks, and yet there is no formal treatment to multiplex structures.

While the grouping of the actors in social networks usually applies some relaxation to the equivalence criterion, in the case of multiplex networks it is desirable to preserve the \emph{multiplicity} of the ties in the network reduction. 
Although it is possible to collapse the different levels into multiplex ties and then apply a global equivalence as with simple networks, significant information gets lost by discarding the multiplicity of the ties. Hence, in order to get a single structure representing a multiple network, we need to combine the distinct particular levels of the relationship, which is feasible by considering the individual perspectives in the modelling.

\subsection{Local role equivalence}
As an alternative to a global equivalence in the reduction of multiple networks, we can apply a \emph{local} perspective in the type of correspondence to be used in the establishment of classes. This means that the standpoints of individual actors are taken separately rather than together in the definition of similarity among the network members in structural terms. In addition, the local equivalences available within social network analysis make it possible to consider not only the primitive relations at different levels, but also the compound relations that go beyond the immediate neighbours of the actor. This is yet another difference compared to the global equivalence types.

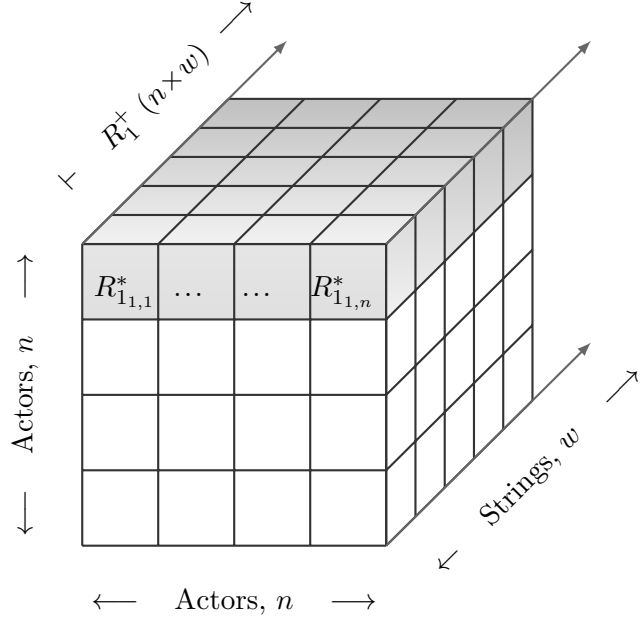
\begin{figure*}
\begin{center}
\begin{tabular}{c}
\begin{minipage}{4cm}
\hspace*{-1cm}
\begin{tikzpicture}[x={(1.0cm,0.0cm)}, y={(0.0cm,1.0cm), z={(0.0cm,0.0cm)}}, scale=1]
    \DrawCoordinateGrid[rotate=180, black!80]{0}{4}{0}{4}{0}{5}
    \DrawCoordinateAxis[rotate=180, thick, black!60]{0}{6}{0}{6}{0}{6}{0}{6}
\node at (-2,-4.75) {$\longleftarrow$ \;\; Actors, $n$ \;\; $\longrightarrow$};
\node[rotate=90] at (-4.75,-2) {$\longleftarrow$ \;\; Actors, $n$ \;\; $\longrightarrow$};
\node[rotate=45] at (2,-3) {$\leftarrow$ \quad Strings, $w$ \quad $\longrightarrow$};
\node[rotate=45] at (-3,2) {$\vdash$  \quad $R^+_{1} ~(n\!\times\!w)$~ $\longrightarrow$};
\node at (-2,-.65) {$R^*_{1_{1,1}}$ \;$\dotss$ \quad $\dotss$ \quad $R^*_{1_{1,n}}$};
\end{tikzpicture}
\end{minipage} \\
\end{tabular}
\setlength{\abovecaptionskip}{20pt}
\caption{Relation Box \emph{with an emphasized relation plane and its role relations}}
\label{fig:RBox}
\end{center}
\end{figure*}

To recognise local equivalences among the actors we rely on a three-dimensional array similar to $\Aa$ where the primitive ties of the network and compounds relations are stacked together. A partially ordered structure by increasing relation similar to the array shown in Figure~\hyperref[fig:RBox]{\ref{fig:RBox}} has been proposed by \citet{WinMan83} for the definition of local equivalences, and they called this device a \emph{Relation-Box}. Figure~\hyperref[fig:RBox]{\ref{fig:RBox}} shows a shadowed horizontal `slice' across the string relations for the outgoing ties of a single actor (the first one in this case) which reflects the actor's activity linked to the rest of the members through the different string relations, both primitives and compounds, that are occurring in the network.


An horizontal slice in the Relation-Box is called a \emph{relation plane}, $R^+_{l}$\!, and encodes the distinct primitive and compound relations that a single actor $l$ has with the rest of the network members. For each network member $l$, there is a vector through the length of the relational plane representing a \emph{role relation}, $R^*_{l_{x,j}}$\!, with actor $j$ and relation $x$. The set of distinct role relations in this case defines the \emph{role set} of actor $l$, and hence there is one role set for each actor from the network that is obtained when the duplicated role relations are removed \citep{WinMan83, WassFau1994}.

A local role equivalence is also a way to characterise social roles in incomplete and in ego-centred networks while preserving the distinction of diverse types of relationships. Besides, \citet{WinMan83} point out that local role equivalence is a generalisation of automorphic equivalence in the sense that both kinds of equivalence involve the same types of role relations. Automorphic equivalence would require not only the same types of role relations, but also the same number of such relations, which implies equal role sets and local role algebras among correspondent actors \citep{Pattison1993}.

Although the Relation-Box theoretically permits consideration of compound relations of infinite lengths, the actors would not be aware of long chains of relations in their surrounding social environment. Thus, based on practical or substantial reasons, it is possible to perform the analysis with a `truncated' version of this array with size $n\times n\times w$, where $w$ is the number of the different primitive and compound ties until a length $k$ that is pre-defined by the researcher.

\subsection{Compositional Equivalence}
\citet{BrePat86} developed one structural correspondence aimed to multiple networks that is based on the individual perspectives of the actors. Although this equivalence type is referred in the literature as `Ego algebra' \citep{WassFau1994}, we call it as \emph{Compositional Equivalence} (CE) since compound relations are taken into account. Thus with CE the analysis of local roles takes the information expressed in the different relation planes of the Relation-Box corresponding to particular network members, and whose rows and columns represent --according to the authors-- the dual structure of the actors and their relations. While such aspect characterizes the local role equivalence type, there is a step forward from a local perspective in the equivalence definition since all the information from particular role relations is generalized to the entire network structure.

The fact that CE generalizes local roles to the entire system implies that this type of correspondence works both at the local and at a `global' level. That is, the establishment of roles and positions in the network are from the perspectives of individual actors, whereas the characterization itself of equivalence is made by considering the relational features that are common to all members in the network. This last feature though works better with middle size networks, and hence CE can be regarded as a local to `middle-range' type of correspondence (Pattison, personal communication).

The local portion of CE lies on the actors' particular views of the system in terms of inclusions among the role relations of the actors' immediate neighbours. Recall that the role relations are recorded in the columns of the individual relational planes, which means that there are in total $n^2$ of these vectors in the network, one for each actor in every relation plane of size $w \times n$ with string relations of length $k$.

Isolated actors in a multiplex network are unable to ``see'' any type of relationships among other actors through the defined links. This implies that role relations for isolates are empty no matter the type of tie or its length, and that any role relations in the relation plane are blank as well. However, connected actors have a different perspective where there is an inclusion among other actors. The collection of inclusions (or lack of them) for each actor or class are reflected in a square array size $n \times n$ called \emph{person hierarchy}  belonging to this entity.

In more formal terms, from the standpoint of a given actor $l$, actor $i$ is `contained within' actor $j$ whenever there is a string $x$ between $l$ and $i$, there is a same type string between $l$ and $j$ 
(cf. \citet[pp. 229]{BrePat86}). Furthermore, the collection of all perceived inclusions in $R^+_l$ represents the person hierarchy $H_l$, which is defined for actors $l, i, j \in \Xx$ and relation $x$ as:
$$
H_{l_{ij}} =
\begin{cases}
1 \quad \text{iff }\; R^*_{l_{xi}} \leq R^*_{l_{xj}} \\
0 \quad \text{iff }\; R^*_{l_{xi}} \nleq R^*_{l_{xj}} \\
0 \quad \text{iff }\; \sum R^*_{l_{xi}} = 0 
\end{cases}
$$
\nov
The last proposition implies that there is no inclusion between actors $i$ and $j$ in the person hierarchy of $l$, and this is either due the lack of containment among these actors or simply because actor $i$ has an empty role set. Notice as well that there is a perceived containment among actors in a given relational plane in case that their role relations are identical, i.e. $H_{l_{ij}} = 1 \Leftrightarrow  R^*_{l_{xi}} = R^*_{l_{xj}}$.

On the other hand, the global part of CE occurs with the union of the different personal hierarchies into a \emph{cumulated} person hierarchy $\Hh$ across actors. This means that $\Hh$ is represented by a single square matrix of size $n \times n$ having the properties of a partially ordered structure, namely reflexive, antisymmetric and transitive. The structural information in the cumulated person hierarchy lays the foundations for categorising the actors and performing a reduction of the network that ---as \citeauthor{BrePat86} pointed out--- comes from the \emph{zeroes} or the absence of inclusions among the different actors.

The partition of the network itself is then a product of a global type of equivalence that is performed on the cumulated person hierarchy. However, we should bear in mind that matrix $\Hh$ does not represent social ties as in $\Aa$, but constitutes a partial order structure indicating the lack of containments among the network members. Hence, we assess classes of actors in the network according to their placement in such graded system that can be visualised through a lattice structure that is aimed for partially ordered sets.

In the next section we illustrate the process of constructing the local and global hierarchies in detail with an application in the reduction of a multiple network. As in \citet{BrePat86}, we study the Florentine families' network classic data set and like the authors we apply CE in the reduction of this social system.

\section{Florentine Families' Network}
\noindent
The Florentine families' network data set \citep{Kent1978, BrePat86, PadAns93} corresponds to a group of people from Florence who had a leading role in the creation of the modern banking system in early 15th century Europe. There are two types of social ties in the network that correspond to Business and Marriage relations among the $16$ prominent Florentine families of which two two stand out for being particularly powerfull and rivals: the \ff{Medici} and the \ff{Strozzi}. The ties of this network are undirected which does not represent any problem for the Marriage ties but it is unfortunate for the Business relations; a circumstance that was remediated by \citeauthor{BrePat86} in their analysis by including measures of power such as families' Wealth and their number of Priorates.\footnote{Data was retrieved from \ucinet}

\begin{figure*}[ht!]
 \centering\includegraphics[width=.86\textwidth]{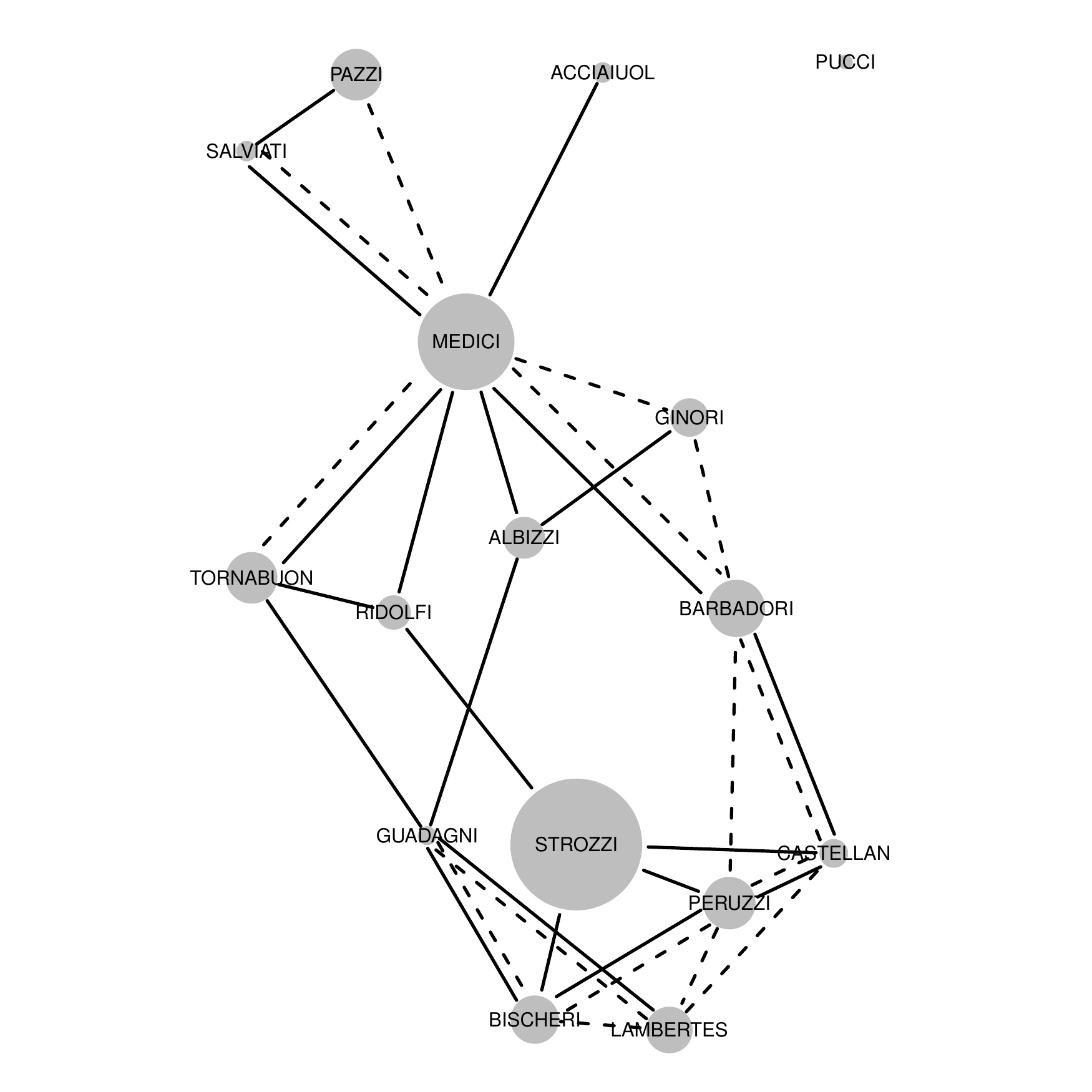}
\setlength{\abovecaptionskip}{20pt}
\caption{Multigraph of the Florentine Families' Network. \emph{Solid edges are Marriage relations, dashed edges are Business ties, and nodes size reflect their financial Wealth}. \emph{Plot made with a force directed layout of the \texttt{multigraph} R package \citep{multigraph17}}}
\label{fig:FFMG}
\end{figure*}

Figure~\hyperref[fig:FFMG]{\ref{fig:FFMG}} depicts the network as a multigraph where different shapes in the edges represent the two kinds of relations. We note in the picture that eight bonds combine Business and Marriage ties in the system and that the network has one component and a single isolated actor represented by the \ff{Pucci} family. A force directed layout algorithm \citep{FrchRein91} has been applied to the graph to avoid crossing edges and also to group together closely related actors. The visualisation gives us initial insights into the general social structure where actors are linked; however, we need to implement some computations in case we want to look at the network relational structure in a form where the different types of tie are interrelated.


%

A crucial part in the modelling of multiple networks is the reduction of the social system as the corresponding relational structure represented by the semigroup is typically large and complex, even for small arrangements. For instance, \citeauthor[][p. 221]{BrePat86} report a semigroup size with an order of $81$ for the Florentine families' network, and this is only considering the two generator relations without attributes. Certainly, it is necessary to work with a more manageable structure in order to obtain better insights in its logic of interlock. 

The reduction of the network implies constructing a relational structure based on a system of roles and positions, which leads to the \emph{role structure} of the network. Thanks to its reduced size, the network role structure is typically a more convenient configuration for a substantial interpretation of the multiple network structure than the `raw' relational arrangement of the system. A key aspect in the creation of the role structure is to preserve the multiplicity of the ties, and we know that local role equivalences allow us to combine different levels in the relationships.

Next, we categorise the actors in the Florentine families' network in terms of CE having actor attributes as generator relations. The first step is to look at the structure product of the actors' views of their neighbours' relations in terms of inclusions, and then we perform the modelling to produce the network positional system to a posteriori analysis of the network role structure.

\subsection{Constructing person hierarchies}
Applying CE in the reduction of a multiple network structure implies the construction of the Relation-Box, which provides the basis of the local part of this type of correspondence. Recall that the Relation-Box is defined by the number of actors in the network and the number of string relations that make up the actors' immediate social ties and eventually the combination of these. Then all inclusions from the individual perspectives are combined into a single matrix that stands for the global part of CE.


\begin{table*}[t]
\caption{Cumulated person hierarchy, $\Hh$, of the Florentine Families' Network of Social Ties, $k=5$. \emph{All computations are made with the \texttt{multiplex} R package \citep{multiplex17}}}
\label{tab:CPH}
\setlength{\belowcaptionskip}{20pt}
\renewcommand{\arraystretch}{1}
\setlength{\tabcolsep}{2.50pt}
\begin{center}
\begin{tabular}{r | c >{$}c<{$} >{$}c<{$} >{$}c<{$} >{$}c<{$} >{$}c<{$} >{$}c<{$} >{$}c<{$} >{$}c<{$} >{$}c<{$} >{$}c<{$} >{$}c<{$} >{$}c<{$} >{$}c<{$} >{$}c<{$} >{$}c<{$} >{$}c<{$}}
\multicolumn{1}{>{$}c<{$} |}{} && 1  & 2  & 3  & 4  & 5  & 6  & 7  & 8  & 9  & 10 & 11 & 12 & 13 & 14 & 15 & 16 \\
\hline\\[-10pt]
1  \ff{Barbadori}                     &&  1   &   1   &   1   &   1   &   1   &   1  &   1   &   1   &   1   &   1   &   0   &   0   &   0   &   0   &   0  &   0   \\
2  \ff{Bischeri}                      &&  1   &   1   &   1   &   1   &   1   &   1  &   1   &   1   &   1   &   1   &   0   &   0   &   0   &   0   &   0  &   0   \\
3  \ff{Castellani}                    &&  1   &   1   &   1   &   1   &   1   &   1  &   1   &   1   &   1   &   1   &   0   &   0   &   0   &   0   &   0  &   0   \\
4  \ff{Guadagni}                      &&  1   &   1   &   1   &   1   &   1   &   1  &   1   &   1   &   1   &   1   &   0   &   0   &   0   &   0   &   0  &   0   \\
5  \scalebox{.90}{\ff{Lamberteschi}}  &&  1   &   1   &   1   &   1   &   1   &   1  &   1   &   1   &   1   &   1   &   0   &   0   &   0   &   0   &   0  &   0   \\
6  \ff{Medici}                        &&  1   &   1   &   1   &   1   &   1   &   1  &   1   &   1   &   1   &   1   &   0   &   0   &   0   &   0   &   0  &   0   \\
7  \ff{Pazzi}                         &&  1   &   1   &   1   &   1   &   1   &   1  &   1   &   1   &   1   &   1   &   0   &   0   &   0   &   0   &   0  &   0   \\
8  \ff{Peruzzi}                       &&  1   &   1   &   1   &   1   &   1   &   1  &   1   &   1   &   1   &   1   &   0   &   0   &   0   &   0   &   0  &   0   \\
9  \ff{Salviati}                      &&  1   &   1   &   1   &   1   &   1   &   1  &   1   &   1   &   1   &   1   &   0   &   0   &   0   &   0   &   0  &   0   \\
10 \scalebox{.93}{\ff{Tornabuoni}}    &&  1   &   1   &   1   &   1   &   1   &   1  &   1   &   1   &   1   &   1   &   0   &   0   &   0   &   0   &   0  &   0   \\
11 \ff{Acciaiuoli}                    &&  1   &   1   &   1   &   1   &   1   &   1  &   1   &   1   &   1   &   1   &   1   &   1   &   1   &   1   &   0  &   0   \\
12 \ff{Albizzi}                       &&  1   &   1   &   1   &   1   &   1   &   1  &   1   &   1   &   1   &   1   &   1   &   1   &   1   &   1   &   0  &   0   \\
13 \ff{Ridolfi}                       &&  1   &   1   &   1   &   1   &   1   &   1  &   1   &   1   &   1   &   1   &   1   &   1   &   1   &   1   &   0  &   0   \\
14 \ff{Strozzi}                       &&  1   &   1   &   1   &   1   &   1   &   1  &   1   &   1   &   1   &   1   &   1   &   1   &   1   &   1   &   0  &   0   \\
15 \ff{Ginori}                        &&  1   &   1   &   1   &   1   &   1   &   1  &   1   &   1   &   1   &   1   &   0   &   0   &   0   &   0   &   1  &   0   \\
16 \ff{Pucci}                         &&  0   &   0   &   0   &   0   &   0   &   0  &   0   &   0   &   0   &   0   &   0   &   0   &   0   &   0   &   0  &   1   
\end{tabular}
\end{center}
\end{table*}

To illustrate the process of constructing person hierarchies we restrict the analysis to the smallest case of the Relation-Box with no compounds that for the Florentine families' network the dimensions are $16\times 16\times 2$. When we look at Fig.~\hyperref[fig:FFMG]{\ref{fig:FFMG}}, we see that apart from \ff{Pucci}, the actor of the network with the lowest number of connections is the \ff{Acciaiuoli} family who is a pendant actor with a single (reciprocated) tie with the \ff{Medici} family. For the direct contacts in the network without compounds, this means that the personal hierarchy of \ff{Acciaiuoli} just includes their immediate neighbour who is the \ff{Medici} family, and hence the only inclusion in the matrix is a reflexive closure corresponding to this neighbour, while all the other possibilities lack containment.

For a two-chain relationship, the person hierarchy of \ff{Acciaiuoli} includes the neighbouring of the \ff{Medici} family as well, i.e. the \ff{Albizzi}, \ff{Barbadori}, \ff{Ginori}, \ff{Pazzi}, \ff{Ridolfi}, \ff{Salviati}, \ff{Tornabuoni}, and in this case the \ff{Acciaiuoli} itself. Note that longer paths include not just the rest of the members in the component, but also those actors who take part in the partial order structures for generators and shorter compounds. Thus for compounds of length $3$, we still see the self-containment for the \ff{Acciaiuoli} family in its person hierarchy.

Each actor $l$ in the network has its own person hierarchy $H_l$ that is based on $R^+_l$ containing the primitive relations and the compounds until a certain length. However, these hierarchies are aggregated into the $\Hh$ --the cumulated person hierarchy-- which is a single matrix of inclusions among all the network members. For the Florentine families' network, the structure of $\Hh$ is represented by the universal matrix\footnote{This is disregarding the isolated actor.} and it makes no differentiation among the actors until it reaches chain of relations of length $4$. It is only from chains of relations with length $5$ or more that $\Hh$ produces a distinction among the actors that is a product of their particular inclusions expressed in $H_l$.

The partial order structure representing $\Hh$ is given in Table~\hyperref[tab:CPH]{\ref{tab:CPH}}, and this set of ordering relations has been reported by \citet[pp. 234]{BrePat86}. This cumulated person hierarchy presents two categories of actors in the network plus the isolated family. One category corresponds to the actors who contain other network members without being contained in them, whereas the other category groups those who are merely contained in other actors without containing them. The partition of this system almost fits the requirements of Structural equivalence, except for the case of the \ff{Ginori} family who is positioned in the same class with the \ff{Acciaiuoli}, \ff{Albizzi}, \ff{Ridolfi} and \ff{Strozzi} even though this actor is not implicated in any inclusions with the rest of the members in this class other than a self-containment.

Therefore, the positional system can have either two classes of collective actors plus the isolated actor, or it can have four classes with pairwise individual positions in the system. Regardless of the option chosen, both reduced arrangements seem to be good representations of the network structure in terms of the patterned social relations, and they serve as the basis for the construction of the role structure of the Florentine families' network. However, a number of attributes from the actors may play a significant part in the establishment of the network positional system, and hence we continue the rest of the analysis of this network by incorporating actor attributes in the establishment of the network role structure.

\subsection{Incorporating family attributes}
The core motivation for this paper is to incorporate in the modelling of the network relational structure significant actor characteristics, which do not have a structural character; that is, traits that are inherent to the actors and do not depend directly on their embedment in the network such as individual centrality measures, dyadic attributes, etc. On the other hand, although actor attributes can be independent variables, they are not ascribed to the actors in the same way as age, gender or other demographic information, but it they are governed by the action of the actors themselves. The belief is that such kind of actor attributes should be part of the modelling of the network positional system and also of the establishment of role structure when the attribute has a structural effect.\footnote{Naturally, extreme cases, e.g. when all or no actors share the attribute, will not have influence on the final structure since they are represented by the identity and the null matrix, respectively.}


\begin{table*}[t]
\caption{Wealth and Number of Priorates of the Florentine Families}
\label{tab:WPFF}
\setlength{\belowcaptionskip}{20pt}
\renewcommand{\arraystretch}{1.0}
\setlength{\tabcolsep}{7pt}
\begin{center}
\begin{tabular}{r | >{$}r<{$} >{$}c<{$} | >{$}r<{$} >{$}c<{$} }
\multicolumn{1}{>{$}c<{$} |}{} & \textsc{Wealth} &  >40  & \text{Number of} & \gtrapprox 34  \\ 
\multicolumn{1}{>{$}c<{$} |}{} & \multicolumn{2}{>{$}c<{$} |}{\footnotesize (\times 1000~\text{Lira})} & \multicolumn{1}{>{$}c<{$} }{\textsc{Priorates}}  & \multicolumn{1}{>{$}c<{$}}{\text{(avg.)}} \\
\hline
\ff{Acciaiuoli}  &  10  &  0  &  53        &   1   \\
\ff{Albizzi}  &  36  &  0  &  65        &   1   \\
\ff{Barbadori}  &  55  &  1  &  \text{NA} &   0    \\
\ff{Bischeri}  &  44  &  1  &  12        &   0   \\
\ff{Castellani}  &  20  &  0  &  22        &   0   \\
\ff{Ginori}  &  32  &  0  &  \text{NA} &   0    \\
\ff{Guadagni}  &   8  &  0  &  21        &   0   \\
\ff{Lamberteschi}  &  42  &  1  &   0        &   0   \\
\ff{Medici}  & 103  &  1  &  53        &   1   \\
\ff{Pazzi}   &  48  &  1  &  \text{NA} &   0    \\
\ff{Peruzzi}  &  49  &  1  &  42        &   1   \\
\ff{Pucci}   &   3  &  0  &   0        &   0   \\
\ff{Ridolfi}  &  27  &  0  &  38        &   1   \\
\ff{Salviati}  &  10  &  0  &  35        &   1   \\
\ff{Strozzi}  & 146  &  1  &  74        &   1   \\
\ff{Tornabuoni}  &  48  &  1  &  \text{NA} &   0   
\end{tabular}
\end{center}
\end{table*}

In the case of the banking families, the power and influence of these families in the 15th century constitute significant characteristics. Table~\hyperref[tab:WPFF]{\ref{tab:WPFF}} gives the Wealth and the number of Priorates of Florentine families as reported in \citet[pp. 744]{WassFau1994}, and these two attribute types, either together or individually, are candidates for the modelling of the network positional system and subsequent role structure. For such type of analysis each attribute is represented with an indexed matrix, and hence reducing the network structure with the actor attributes resembles the process we just applied to the marriage and business relations with CE, except that now there are additional generators to the social ties representing the attributes.

We note in Table~\hyperref[tab:WPFF]{\ref{tab:WPFF}} that each category has two columns, one for the absolute values and another that marks the limits of these values according to a cut-off value. In one case we differentiate the very wealthy families from the ``modestly'' rich actors in the network by adopting a cut-off value of $40000$ Lira, which approximates the average of their financial resources.\footnote{Actually, the mean is $42.56$, and the \ff{Lamberteschi} family lies in this limit, but rounding the cut-off value to $40$ makes more sense for the analysis.}  On the other hand, as regards the number of Priorates, it seems reasonable to assume that the lack of information implies that these actors did not have a large number of jurisdictions at that time, if at all, and the cut-off lies in the average of the accessible number of priorates that is rounded to $34$. As a result, there are two vectors of binary values that make the diagonal of the indexed matrices representing the actor attributes, which are additional generators for constructing the network relational structure.

We continue the analysis of the banking network by applying CE for grouping the actors in the construction of the positional system with actor attributes. The difference is that now the Relational-Box on which the person hierarchies are based includes the additional generators representing the attribute-based information. Since indexed matrices just have information on their diagonals, it means that the different person hierarchies in the network include self-containments whenever the actor has the attribute. For example, while the person hierarchy of \ff{Acciaiuoli} for immediate ties comprises just the \ff{Medici}, with actor attributes it will include the \ff{Acciaiuoli} family itself when $k=1$ because this particular actor is politically very powerful with a number of priorates larger than the average. Naturally, the rest of the actors in the network will follow the same logic, and the arrangement of the cumulated person hierarchy will be affected by the different personal views on inclusions, which are restructured due to the presence of actor attributes.

\begin{figure*}[ph!]
\setlength{\tabcolsep}{2pt}
\begin{tabular}{l}
	    \includegraphics[width=0.52\textwidth]{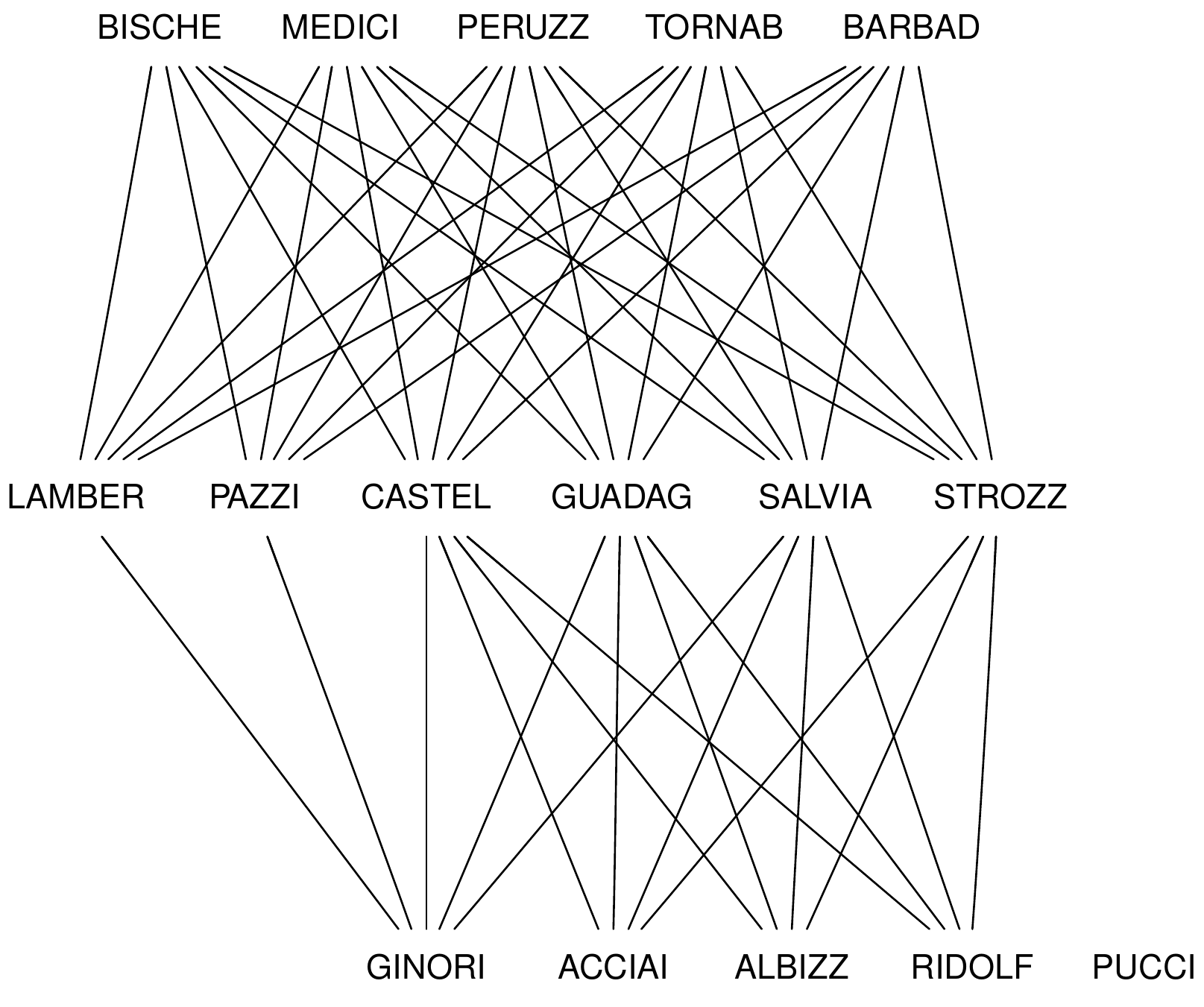} \\[-10pt]
\hspace{6cm}\includegraphics[width=0.52\textwidth]{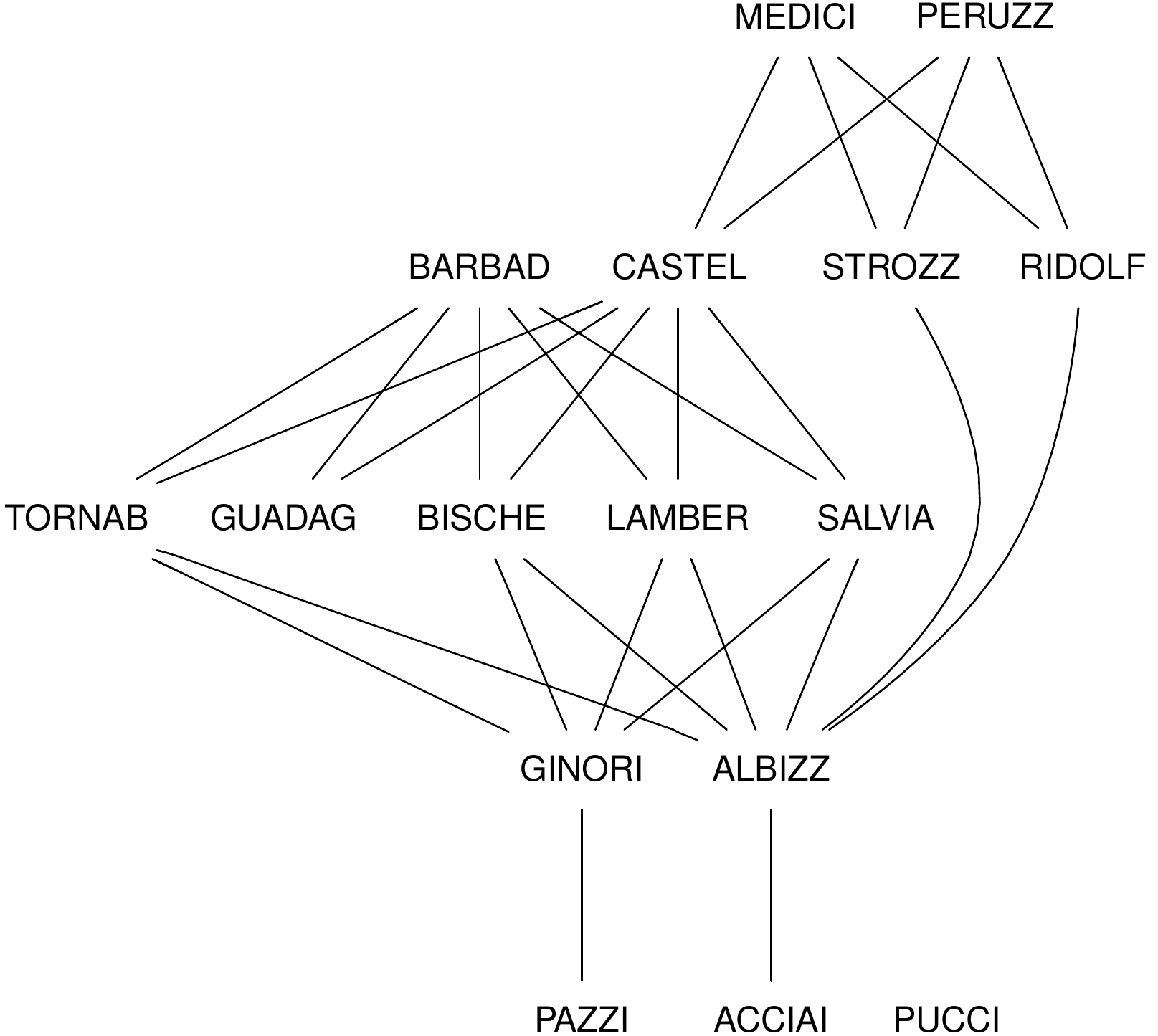} \\[-10pt]
	    \includegraphics[width=0.52\textwidth]{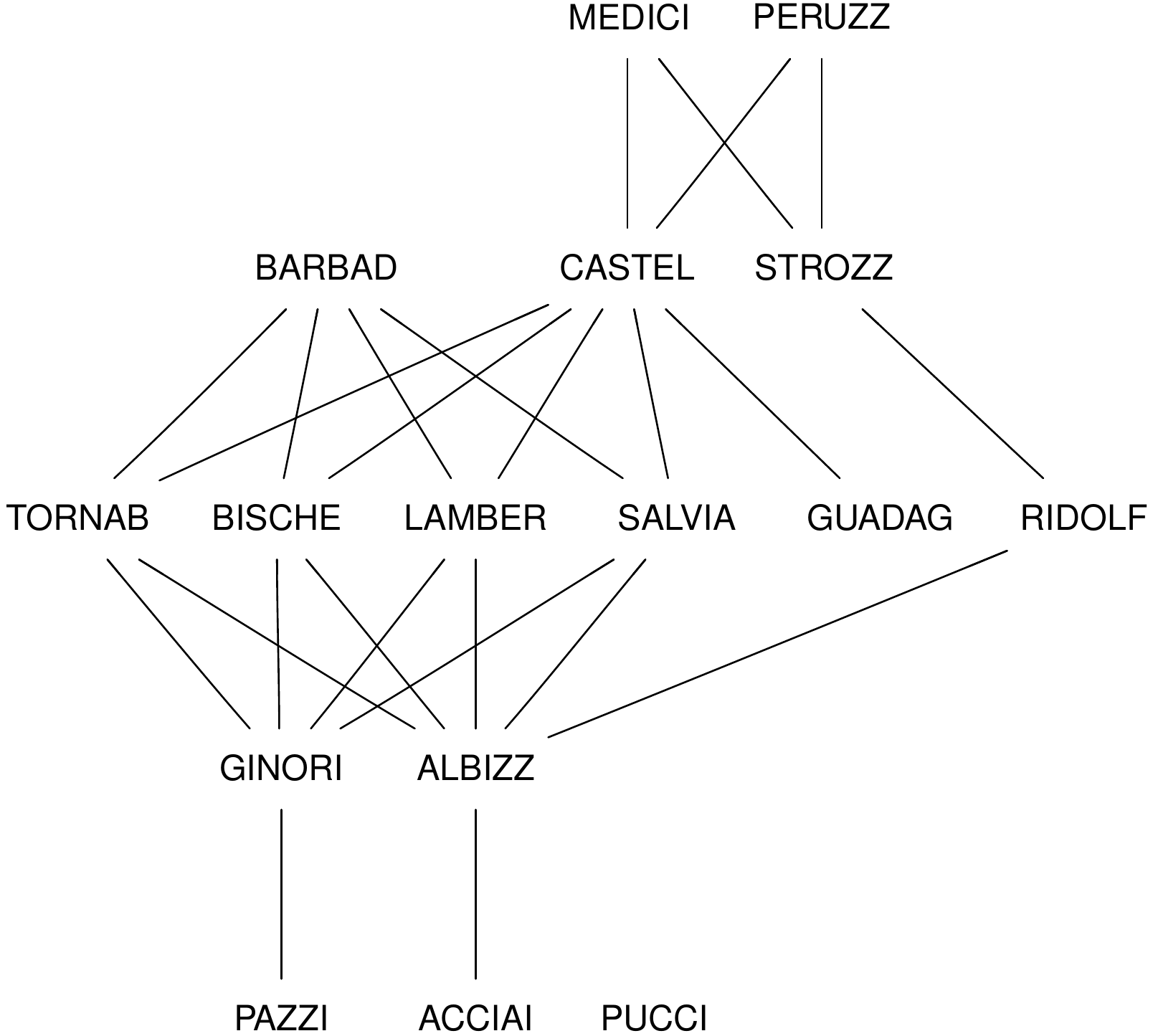} \\
\end{tabular}
\setlength{\abovecaptionskip}{20pt}
\caption{Hasse Diagrams of $\Hh$ for the Florentine Banking Network with Actor Attributes. \emph{Top to bottom: with Wealth, $k=5$; with number of Priorates, $k=4$; with Wealth \& Priorates, $k=4$. Plots made with the \texttt{multiplex} \citep{multiplex17} and the \texttt{Rgraphviz} packages \citep{Rgraphviz16}}}
\label{fig:HDFFA}
\end{figure*}

Figure~\hyperref[fig:HDFFA]{\ref{fig:HDFFA}} shows $\Hh$ in a graphic mode for the banking network with Business and Marriage ties together with Wealth, number of Priorates, and both actor attributes combined. These pictures are lattices known as \emph{Hasse diagrams}, which depict the inclusion levels in the hierarchy where the lower bound elements are contained in the upper bound elements whenever there is a link among them. For instance, in each diagram the inclusion ties of the \ff{Medici} family contain the inclusion ties of the \ff{Acciaiuoli} and the \ff{Pazzi} families, whereas in any of the cases there is a containment relation among these last two actors.

It is important to note, however, that although the different levels in the Hasse diagrams try to reflect the ranks in the partial order structures, there can be ambiguities in the placements depending on the diagram structure. For example, the \ff{Guadagni} family is always placed in the most intermediary level of the diagrams in Fig.~\hyperref[fig:HDFFA]{\ref{fig:HDFFA}}, but in a couple of cases this actor does not contain any other actor in $\Hh$. Likewise, the inclusion ties of \ff{Barbadori} contain the ties of other actors while it is not being contained at all similar to \ff{Medici} and \ff{Peruzzi}, and it may be best depicted at the same level with these actors. Such aspects deal with aesthetics rather than the structural representation of the partially ordered system, however.

All partial orders shown are emerging structures with the smallest value of $k$. This means that there are `zeroes' among connected actors in $\Hh$ with compounds of such lengths, which allows us to rank classes of actors according to the CE criteria. In the case of the actors' wealth, the structure of $\Hh$ remains unaltered after compounds of length $5$, but in the other two cases the cumulated person hierarchies involve a lower number of inclusions with larger $k$. However, shorter chains of relations imply more truthful individual viewpoints than ordered systems with longer compounds and they are therefore preferred.

These diagrams very clearly show that the attributes of the actors such as their monetary wealth and political power have an impact on the relational structure of this particular network. If we look at the diagrams in Fig.~\hyperref[fig:HDFFA]{\ref{fig:HDFFA}}, we note that there is a further differentiation in the network in all three cases when considering actor attributes in the modelling. Apart from the isolated actors, whose personal hierarchy corresponds to the null matrix, the cumulated person hierarchy for Wealth clearly involves three levels, whereas there are five levels in the diagrams for the number of Priorates, and for the two attributes together.

As a result, the positional system with Wealth differentiates three categories of actors plus the isolated node where the largest class in the previous classification is now divided into two categories. Thus the personal wealth has a structuring influence in the network, and this makes a lot of sense; the richest actor of the banking network is the \ff{Strozzi} family who is no longer in the same class as the \ff{Acciaiuoli}, \ff{Albizzi}, \ff{Ridolfi} families, but is placed in another category with other actors having much more social and financial capital.

When we look at the Number of Priorates there is even more differentiation in $\Hh$ than we saw when just considering the Wealth of the actors. Apart from the families who contain most of the network members, i.e. the actors `at the top' (\ff{Medici}, \ff{Peruzzi}, \ff{Barbadori}), and conversely the actors `at the bottom' (\ff{Pazzi}, \ff{Acciaiuoli}, \ff{Guadagni}) who are contained in the rest of the network component, there is ambiguity with the rest of the actors and they can be classed in different ways. We get a similar picture when both attributes are taken together (cf. Fig.~\hyperref[fig:HDFFA]{\ref{fig:HDFFA}}c), where the `top' and `bottom' actors in the diagram representing $\Hh$ remain unambiguously placed, whereas the categories of the actors in-between require interpretation.

Theory can guide us in the establishment of the categories in the positional system in the two last cases. We also need to determine which of the resulting role structures product of the positional system provides the best insights into the relational interlock of the multiple network structure. Such aspect constitutes one of the last steps in the modelling of the system and we look at the reduced relational structures of the banking families' network.


\subsection{Positional System of the Florentine families' network}
The main challenge in establishing the positional system of the network is to find the sets of collective relations that produce the most meaningful network role structure. That is, a reduced system that provides an insight into the logic of interlock of the network relations, and this is typically achieved with the role structure having the smallest possible dimension. The logic of interlock is a kind of rationality that is shaped by different algebraic constraints expressed in the final relational structure where the different types of ties and the relevant actor attributes are interrelated in this case.

Although the class membership with the Wealth attribute with three defined classes of collective actors seems straightforward, there are ambiguities both as regards the amount of Priorates and when the two features are combined. Such uncertainties arise because a number of actors in the network can be classed in different ways according to their respective locations in the partial order structures of $\Hh$, and for the time being we concentrate our analysis on the two cases where political power is involved. Hence, assuming that the isolated actor of the network makes its own class, we need to categorise the eight actors that are neither at the `top' nor at `bottom' of the hierarchies shown in Figs.~\hyperref[fig:HDFFA]{\ref{fig:HDFFA}} and~\hyperref[fig:HDFFA]{\ref{fig:HDFFA}}c, and in both arrangements the placement of the actors at the different levels aims to reflect the set of containments in the partial order structures with an aesthetical representation in the lattice.\footnote{That is why \ff{Barbadori} and \ff{Guadagni}, for example, who are unequivocally part of the same class as the top and bottom actors, respectively, are located at intermediary levels in the diagram.}

Now we look closer at the in-between actors in the two hierarchies where political power is involved. From Table~\hyperref[tab:WPFF]{\ref{tab:WPFF}} we obtain the assignment of these families with respect to the two attributes, and the next upper and lower vectors give the categories for Wealth and number of Priorates, respectively: 

\medbreak
\begin{strip}
\renewcommand{\arraystretch}{.8}\setlength{\tabcolsep}{4pt}
\hspace{100pt}
\begin{tabular}{ccccccccc}
\ff{Albizz} & \ff{Bische} & \ff{Castel} & \ff{Ginori} & \ff{Lamber} & \ff{Ridolf} & \ff{Salvia} & \ff{Strozz} & \ff{Tornab}  \\
     0 &      1 &      0 &      0 &      1 &      0 &      0 &      1 &      1  \\
     1 &      0 &      0 &      0 &      0 &      1 &      1 &      1 &      0  
\end{tabular}
\end{strip}

\nov
Certainly, one possibility is that all these actors are grouped together onto a single class irrespective of their economic or political power, and in this way we have a positional system with three categories of collective actors for both Priorates, and also for Wealth and Priorates. The arrangements of roles for Business and Marriage are then equal and all the positions are represented by actors who are both very wealthy and powerful in political terms (of course disregarding \ff{Pucci}). This means that the two attribute types are represented in the positional system by identity matrices with no structuring effect in the system of roles. In order to have an effect from the Wealth and the number of Priorates on the role structure, we need to make a differentiation between classes of actors with respect to these attributes, and this is only possible by having characteristic strings not acting as neutral elements in the construction of the semigroup of relations.

A straightforward way to achieve a structuring effect of diagonal matrices is by separating the actors with `ones' in the intermediate category from the actors with `zeroes' in the vector corresponding to this attribute type. Hence we end up with a positional system that has four categories of collective actors, and for the number of Priorates, for instance (the second row above), then \ff{Bischeri}, \ff{Castellani}, \ff{Ginori}, \ff{Lamberteschi} and \ff{Tornabuoni} will make their own class. This means that the attribute string is no longer represented by an identity matrix and the semigroup of the role structures for Business, Marriage, and number of Priorates will record different compounds of social roles with class attributes. However, the role structure for Priorates (not shown here) results being relatively large and complex.

Conversely, if we model the network relational system with both attributes at the same time, we first differentiate the \ff{Strozzi} who is a very powerful family both politically and economically, and second we differentiate \ff{Castellani} and \ff{Ginori} who are actors who are neither very wealthy nor have much political power. By grouping the last two actors into a single class we again avoid having the identity matrix, and the role structure of the network in this case has fewer representative strings, which means that we expect a more tractable substantial interpretation of the role interlock than when just considering the Priorates. The fact that the role structure gets smaller rather than larger as one would expect with another generator is because the two social roles and both class attributes are equated, and the relational structure of the positional system is then based just on two generators. When we equate roles or attributes we get a poorly informative role structure where we need to interpolate the roles and collective characteristics in the analysis.\footnote{Besides, assigning \ff{Strozzi} in the central class does not affect the role structure at all.}

A third possibility is to combine the Business and Marriage ties with Wealth in the analysis, in which case the class system of actors takes the levels given in the Hasse diagram of Fig.~\hyperref[fig:HDFFA]{\ref{fig:HDFFA}}a. The positional system in this case implies that the Marriage ties do not follow a particular pattern in the role structure, whereas Business ties and Wealth role relations follow a core-periphery structure as the matrices below show it:

\begin{center}
\begin{tabular}{ccc}
\begin{tabular}{>{$}c<{$} >{$}c<{$} >{$}c<{$} }
	 1 & 1 & 1 \\
	 1 & 1 & 0 \\
	 1 & 0 & 0 \\
\end{tabular}
&
\begin{tabular}{>{$}c<{$} >{$}c<{$} >{$}c<{$} }
	 1 & 1 & 1 \\
	 1 & 1 & 1 \\
	 1 & 1 & 1 \\
\end{tabular}
&
\begin{tabular}{>{$}c<{$} >{$}c<{$} >{$}c<{$} }
	 1 & 0 & 0 \\
	 0 & 1 & 0 \\
	 0 & 0 & 0 \\
\end{tabular} \\[20pt]
Business & Marriage & Wealth \\
\end{tabular}
\end{center}

There are no ambiguities in the categorisation of actors in the banking network with the financial Wealth of the actors, which leads to a univocally substantial interpretation of the role structure for this positional system. However, the main advantage with these generators is that the role structure gets smaller than with the previous two settings, allowing a more transparent interpretation of the role interlock, even though we are aware that a different logic may arise in the role structure when considering the number of Priorates. The reader can refer to \citet{Ostoic2018} for an extended analysis of the role structure and role interlock of this particular network.

\section*{Discussion}
\noindent
The structuring effect of attribute-based information in the reduction of multiplex networks constituted the most significant aspect covered in this paper where one of the main challenges has been preserving the multiplicity of the different types of tie. In this sense, the notion of Compositional Equivalence defined by \citeauthor{BrePat86} allows us to reduce the network structure without dropping the relational differentiation, and we extend the positional analysis to non-ascribed characteristics of the actors in the network, which are included as generator relations in the form of diagonal matrices. There is a strong belief that attribute-based information enriches the substantial interpretation of the relational structure of the network, and this is so irrespective of whether the relational system is in a reduced or in a full format.

Even though the reduction of the network can bring some ambiguities, aggregated structures are more manageable for substantial interpretation of the relational logic in multiple network structures that are complex systems by definition. CE has proven to be a valuable option for mid-sized networks; however, theoretical guidance is required both for the selection of the attribute types and for the establishment of the positional system and subsequent role structure.

There still some important aspects that need to be accounted. The first concern deals with directed multiplex networks, in which the application of CE typically requires counting with relational contrast reflected in the transposes of the ties. A second aspect is the rationale behind relational structures, which is expressed by algebraic constraints governing the system, including sets of equations among strings, hierarchy in the relations, and interrelations between the different types of tie occurring in the network. These aspects are barely mentioned here and their treatment is out of the scope. Finally, a statistical approach to the modelling is required for larger network structures and statistical methods for multiplex networks can serve to complement the modelling process either in an early stage of the analysis or by providing relational and role structures having both fixed and random effects with attributes.

\medbreak

  \theendnotes

\bigbreak

\setstretch{0.86}
  \bibliography{CEAAarXiv}

\newpage
\clearpage

\end{document}